\documentclass{article}

\begin{document}

\begin{center}
{\Large \textbf{A very singular property}\footnote{Originally published in Italian on the popular science magazine \emph{Darwin}, n. 34, November/December 2009.}}\\
By using the \emph{Fermi} satellite, a group of physicists discovers a new type of gamma-ray active nucleus\\
\vskip 12pt
\emph{Luigi Foschini\\
INAF - Osservatorio Astronomico di Brera}\\
\normalsize
\end{center}

\emph{Active galactic nuclei (AGNs)} are galaxies with a million, or even billion, solar masses spacetime singularity (commonly known as \emph{black hole}) in their center. The strong gravitational field of the central singularity modifies the nearby environment, attracting gaseous matter, which in turn structures itself to form an equatorial \emph{accretion disk}. The dimensions of this disk depends on the mass of the central singularity: as an example, a 100 million solar masses singularity can have a disk with radius equal to one hundredth of parsec (1 parsec = 3.26 light years). The sphere of gravitational influence is greater and can affect the motion of stars distant up to about tens of parsecs. The matter falling onto the singularity converts the gravitational energy into optical and ultraviolet radiation, which can be observed from Earth. Therefore, the galaxy appears with a particularly bright nucleus and hence it is named AGN. 

But the accretion disk is not the only consequence of the presence of a spacetime singularity. A \emph{corona} is formed above and below the disk, as around the Sun, which is a kind of atmosphere reaching the temperature of millions of degrees and emitting radiation in the X-ray energy band. The radiation emitted by the disk ionizes the gas outside the disk itself, forming a plasma, which in turn emits photons at specific wavelengths (emission lines), depending on the chemical elements composing it (hydrogen, oxygen, carbon, and so on). These emission lines are affected by the central gravitational field and their profile results to be broadened. By measuring the width of the line profile, it is possible to calculate the orbital velocity of the plasma and thus estimate the mass of the singularity. 

The plasma tends to place itself into two main regions at different distances from the black hole: the nearest (about $0.1-1$ parsecs) is named \emph{broad-line region} (BLR), since the plasma have to move at high velocity (greater than 2000 km/s and up to a few tens of thousands of kilometers per second) to maintain the orbit at this distance from the singularity and this velocity determines a significant broadening of the emission lines profile. The second region is more distant from the singularity (about $10$ parsecs and more) and, therefore, it requires smaller orbital velocity (a few hundreds of kilometers per second), which in turn means a smaller broadening of the line profile. Thus, this region is named \emph{narrow-line region} (NLR). 

In the middle of these two regions, at about $1-10$ parsecs from the singularity, there is a toroidal zone placed on the equatorial plane and made of molecules of hydrogen and cosmic dust. If one observes the AGN through the equatorial plane, this \emph{molecular torus} can obscure the direct view of the nucleus. 

In addition to the reorganization of the parsec-scale nearby environment, with particular structures like the accretion disk, the corona, the BLR, the torus, and the NLR, the spacetime singularity is also able to eject plasma at very high velocities, close to that of light. About 10\% of AGNs develops these \emph{relativistic jets}, which forms close to the polar regions and extends for several thousands of parsecs and, sometimes, up to millions of parsecs in the intergalactic space. In the remaining 90\% of cases, the AGNs develop \emph{outflows} that cannot collimate into a jet, but can reach velocities of about a few tenths of the light speed. 

This structure of AGNs is the result of a plethora of researches culminated during nineties with the proposal of the \emph{unified model} by Robert Antonucci\footnote{R. Antonucci, 1993, ARA\&A, 31, 473.}, C. Megan Urry and Paolo Padovani\footnote{C.M. Urry \& P. Padovani, 1995, PASP, 107, 803.}. According to this model, it is possible to account for all the observational differences by considering a single structure, but viewed at different angles. Therefore, AGNs without relativistic jets observed along the equatorial plane -- and hence obscured by the torus -- are named Type 2 Seyfert galaxies (shortly, Seyfert 2). This class of AGNs is characterized by an optical spectrum with only narrow lines emitted by the NLR, since the obscuring torus does not allow a direct view of the BLR. The Type 1 Seyfert galaxies are those observed at intermediate angles between the equatorial plane and the pole, without the torus obscuring the view of the nucleus. In this case, the optical spectrum shows the lines from both the BLR and the NLR. 

A peculiar class of AGNs is that of Narrow-Line Seyfert 1 (NLS1): in this case, the viewing direction is along the pole (hence no obscuration), but the optical emission lines are anomalously narrow, although they are emitted by the BLR, indicating an orbital speed smaller than 2000 km/s. This phenomenon can be explained in two ways, although none of them seems to be presently conclusive. In one theory, elaborated by a group of Italian researchers led by Roberto Decarli\footnote{R. Decarli et al., 2008, MNRAS, 386, L15.}, the BLR has a disk-like shape (instead of shell-like as in classical Seyferts) and, then, by observing it pole-on, we can see the orbital motion of the plasma, but there is no component along the viewing direction (this causes the broadening of the lines) and therefore the lines have profiles narrower than usual. A second hypothesis, developed by an international team led by Alessandro Marconi\footnote{A. Marconi et al., 2008, ApJ, 678, 693.}, is that this type of source has a disk with a very high accretion rate, thus causing the radiation pressure to push the BLR farther from the central singularity, where a smaller velocity is needed to maintain the orbit (and therefore the lines are narrower). 

When a relativistic jet is present, then this type of AGNs is called radio galaxy. However, if the viewing angle is very close to the jet axis, then the effects of the special relativity becomes dominant and these AGNs are named blazars. This term derives from the contraction of the words BL Lac and quasar. Indeed, at the end of seventies, there were two classes of objects with similar characteristics: flat-spectrum radio quasars (FSRQ) and BL Lac objects, which were so called because of the peculiar variable star\footnote{Only at the end of seventies, it was possible to understand the extragalactic nature of BL Lac.} (BL) in the constellation of Lacerta. During a renowned conference held in Pittsburgh in 1978, Ed Spiegel proposed that these sources, sharing common observational characteristics, were collected in the same class to be named blazar.

These episodes teach us that it is not sufficient to observe anything to immediately understand its nature. The evolution of the knowledge is made of little steps, often uncertain because based on partial observations, which have to be assembled like in a puzzle. The cosmic sources emit a vast continuous set of radiations and particles, while human beings need of instruments to detect them, which have necessarily limited and finite characteristics. The most used information messenger is the photon, which is the quantum of the electromagnetic field, although during the last years several types of advanced instruments have been developed to detect other information carriers, like neutrinos, cosmic rays, and gravitational waves. 

The electromagnetic radiation can be divided into different domains, depending on the photon energy: from the lowest energies of radio frequencies to the highest of $\gamma$ rays, passing through infrared, optical, ultraviolet and X-ray. An individual detector able to measure signals all over the electromagnetic spectrum does not exist (at least we do not know -- yet? -- how to build it) and, therefore, we have to adopt different techniques to detect photons at different energies, also because they interact in different ways with the matter. 

Our cosmic view has always been influenced by the instruments used to observe, because each of them collect only the information within a restricted energy or frequency band. For example, speaking of blazars, the first radio surveys made during fifties and sixties, with rough instruments, with sensitivity enough to detect the brightest sources, discovered more FSRQ than BL Lac. Instead, during seventies, with the advent of the first X-ray satellites, the opposite occurred. Indeed, FSRQ have stronger radio emission than BL Lac, while the latter are brighter in X-rays. Today, this selection effect is dimming, since the sensitivities of the instruments are increasing. However, it was common to speak about radio-selected and X-ray selected blazars till mid-nineties, when P. Padovani and P. Giommi\footnote{P. Padovani \& P. Giommi, 1995, ApJ, 444, 567.} understood that both types of sources belong to the same class. To realize this, they collected all the available information at each frequency. That is, the only way to minimize this instrumental selection bias, is to gather all the observations performed with different instruments. 

At the end of nineties, a group of Italian researchers\footnote{G. Fossati, L. Maraschi, A. Celotti, A. Comastri \& G. Ghisellini, 1998, MNRAS, 299, 433}$^,$\footnote{G. Ghisellini, A. Celotti, G. Fossati, L. Maraschi \& A. Comastri, 1998, MNRAS, 301, 451.} gave a scientific basis to the intuition of Spiegel, by unifying the blazars into a \emph{sequence}, where the shape of the broad-band emitted electromagnetic spectrum depends on the luminosity of the source. At one extreme of the sequence there are the quasars, which are the most luminous sources with the maximum of their emission in the MeV-GeV $\gamma-$ray energy band, while at the opposite extreme, the lowest luminosity blazars (BL Lac objects) have the peak of their emission still at $\gamma$ rays, but at TeV energies.

In other words, as outlined above, the AGN unified model interprets the observations in term of the same structure as observed by different viewing angles, so that we see some effects and not others. To summarize, we started from a zoo of sources, based not only on the viewing angle, but also on the type of instrument used to discover them and, step-by-step we unified all of them in a single model. 

However, this theory is not conclusive and many aspects of the unified model are not yet understood. Just as example, it is not known how jets are formed and why do they form only in a small part of AGNs, but not in all. One thing was noted: while jets develop in elliptical galaxies, AGNs without jets are hosted by spirals. This suggested that there should be a link between the relativistic jet and the host galaxy. But today, a recent discovery makes this paradigm to waver.

Indeed, around 2000, the radio emission typical of relativistic jets was discovered to be present in some NLS1s, which are generally without jets. This alone does not mean anything, because radio emission has been detected also in some Seyferts and it is thought it can come from the basis of a not yet developed (aborted?) jet. The conclusive proof to confirm the existence of a relativistic jet is the detection of high-energy $\gamma$ rays. The reason is due to the fact that photons with energies greater than about 1 MeV tends to create electron-positron pairs as soon as they interact with a third particle or field (generally another photon), which allows them to fulfill the momentum conservation law. Since we cannot detect electrons and positrons so far, the information is lost. If photons are traveling at relativistic speeds, this strongly reduces the probability of interaction with other photons and, then, they can escape from the jet and reach us. Thus, if we observe high-energy $\gamma-$ray photons from a cosmic source, this means that relativistic effects -- like those in a jet -- are at work. 

The proof that NLS1s host effectively a relativistic jet arrived this year, thanks to the \emph{Fermi} satellite that discovered high-energy ($E>100$ MeV) $\gamma$ rays from a NLS1 named PMN J0948+0022, distant 5.5 billions of years and apparently placed in the Sextant constellation\footnote{A.A. Abdo et al., 2009, ApJ, 699, 976.}$^,$\footnote{L. Foschini et al., 2009, in Proc. Conf. on ``Accretion and Ejection in AGN: a Global View'', ASP Conf. Proc., ed. L. Maraschi, G. Ghisellini, R. Della Ceca \& F. Tavecchio (San Francisco, CA: ASP), in press (\texttt{arXiv:0908.3313}).}. 

The satellite for astrophysics \emph{GLAST} (\emph{Gamma-ray Large Area Space Telescope}) was launched by NASA on June 11, 2008, and was immediately renamed \emph{Fermi Gamma-ray Space Telescope} (hereafter \emph{Fermi}) to honor the Italian physicist Enrico Fermi, Nobel Prize for physics in 1938 for his studies on neutron-induced radioactivity, but who also made important discoveries in astrophysics, like, for example, the mechanisms for particle acceleration that are still today named after him. The \emph{Fermi} satellite carries two instruments onboard: one for the Gamma-Ray Bursts (GRB) is named Gamma-ray Burst Monitor (GBM); the other is the Large Area Telescope (LAT). The former is an all-sky monitor operating in the 8 keV - 40 MeV energy range, without imaging capabilities. The latter is the main instrument, can produce images in the range $0.1-300$ GeV with a few arcminutes of angular resolution (at 1 GeV) and has a large field of view (2.4 sr), which allowed it to perform an all-sky every 3 hours (2 orbits). This telescope was built with a significant Italian contribution, both financial (Agenzia Spaziale Italiana, ASI) and hardware (Istituto Nazionale di Fisica Nucleare, INFN). The Istituto Nazionale di Astrofisica (INAF) collaborates today to the scientific data analysis\footnote{Obviously the LAT Collaboration benefitted of the work of several institutes from USA, Europe and Japan (see \texttt{https://www-glast.stanford.edu/institutions.html}). However, being the original article for an Italian popular science magazine, I have focused on the Italian contribution.}. 

The discovery of high-energy $\gamma$ rays from PMN J0948+0022 has been followed by other detections of NLS1s and these results are now in press\footnote{A.A. Abdo et al., 2009, ApJ Letters, accepted for publication (\texttt{arXiv:0911.3485}).}.

The importance of these results has to be evaluated on two different points of view. The first is that a new class of $\gamma-$ray AGNs joins the well-known classes of blazars and radio galaxies. Although these NLS1s display a behavior\footnote{As shown by the multiwavelength campaign performed  on PMN J0948+0022 between March and July 2009. See A.A. Abdo et al., 2009, ApJ, 707, 727.} similar to blazars, with a fully developed relativistic jet, they are \emph{not} blazars, as it was already known from the study of their optical spectrum, which shows a BLR different from that in blazars. However, this seems to be not important with respect to the generation of a relativistic jet, where instead it is important the energy density developed in the BLR and the angle with which the jet sees the radiation coming from the BLR. 

The second point, perhaps the most important, is the fact that NLS1s are hosted in spiral galaxies -- not ellipticals as for blazars and radio galaxies -- and, therefore, the paradigm linking the relativistic jet with the host galaxy breaks down. 

The jet is then something depending only on the central spacetime singularity, although we do not know yet what are the factors determining its generation. The best candidates are the mass, the disk accretion rate and the spin of the singularity, but perhaps something is missing or some link has not yet understood, because there is no answer yet to the question: how and why relativistic jets are formed?

The objections raised against this discovery are focused on the hopeless attempts to reconcile the characteristics of this new class of sources to those known, i.e. blazars and radio galaxies. The same occurred when NLS1s were first discovered in the middle of eighties by D. Osterbrock and R. Pogge\footnote{D.E. Osterbrock \& R.W. Pogge, 1985, ApJ, 297, 166.}: at the beginning, they were thought to be a simple subclass of Seyferts and, only after several years, it was evident that NLS1s were a true class\footnote{R.W. Pogge, 2000, New Astronomy Reviews, 44, 381.}. 

Today, NLS1s enter in the restricted group of AGNs with relativistic jets and history repeating: there is someone who thinks they are a subclass of blazars or just radio galaxies. In the first case, as already outlined, the optical spectrum indicates that it is not so and no blazars with a similar optical spectrum do exist. In the second case, the difference is even more striking: in addition to the different optical spectrum, NLS1s emit a greater power and are lacking of extended radio structures, which instead are typical of radio galaxies and can reach sizes up to Mpc. NLS1s are very compact with at maximum parsec scale structures, which in turn are present also in blazars. In both cases, there is the fundamental difference of the spiral host galaxy, which irreparably smashes the paradigm of the link ellipticals-jets. 

To conclude, independently of the point of view, an anomaly is present and, therefore, it is worth studying. Researchers are actively searching for new $\gamma-$ray detections of NLS1s and organizing observing campaign on the sources now known. The discovery of $\gamma$ rays from NLS1s offers a lot of stuff worth thinking and opens new horizons in this research field.

\end{document}